\def\ifarxiv{\iftrue}     

\ifarxiv
\documentclass[12pt,a4paper]{article}

\usepackage{amsmath}

\makeatletter
\let\old@makecaption=\@makecaption
\def\@makecaption{\small\old@makecaption}
\makeatother

\makeatletter
\let\old@startsection=\@startsection
\renewcommand{\@startsection}[6]{\old@startsection{#1}{#2}{#3}{#4}{#5}{#6\mathversion{bold}}}
\makeatother

\else
\documentclass{elsart4-1}

\usepackage[english,francais]{babel}

\def\og{\leavevmode\raise.3ex\hbox{$\scriptscriptstyle\langle\!\langle$~}}
\def\fg{\leavevmode\raise.3ex\hbox{~$\!\scriptscriptstyle\,\rangle\!\rangle$}}

\fi

\usepackage{graphicx}
\usepackage{amssymb}

\ifx\genfrac\sdflkaj

\else

\fi
\newcommand{\sfrac}[2]{{\textstyle\frac{#1}{#2}}}
\ifx\half\sdflkaj
\newcommand{\half}{\sfrac{1}{2}}
\else
\renewcommand{\half}{\sfrac{1}{2}}
\fi

\newcommand{\lrbrk}[1]{\left(#1\right)}

\newcommand{\vev}[1]{\langle#1\rangle}

\newcommand{\comm}[2]{[#1,#2]}

\newcommand{\state}[1]{\mathopen{|}#1\mathclose{\rangle}}
\newcommand{\set}[1]{\{#1\}}

\newcommand{\algr}[1]{\mathfrak{#1}}

\newcommand{\alSU}{\algr{su}}

\newcommand{\alSO}{\algr{so}}

\newcommand{\alPSU}{\algr{psu}}

\def\<{\begin{eqnarray*}}
\def\>{\end{eqnarray*}}

\newcommand{\pint}{\makebox[0pt][l]{\hspace{2.8pt}$-$}\int}

\newcommand{\superN}{\mathcal{N}}

\newcommand{\fld}{\mathcal{W}}
\newcommand{\fldZ}{\mathcal{Z}}
\newcommand{\fldF}{\mathcal{F}}
\newcommand{\fldX}{\mathcal{X}}
\newcommand{\cder}{\mathcal{D}}
\newcommand{\Op}{\mathcal{O}}
\newcommand{\Tr}{\mathop{\mathrm{Tr}}}

\newcommand{\dil}{\mathfrak{D}}
\newcommand{\parity}{\mathfrak{p}}
\newcommand{\genJ}{\mathfrak{J}}
\newcommand{\gen}[1][J]{\mathfrak{#1}}
\newcommand{\charge}{\mathcal{Q}}
\newcommand{\contour}{\mathcal{C}}

\ifx\href\asklfhas\newcommand{\href}[2]{#2}\fi
\newcommand{\arxivno}[1]{\href{http://arxiv.org/abs/#1}{#1}}

\begin{document}

\ifarxiv

\setcounter{page}{0}\thispagestyle{empty}
\begin{flushright}\footnotesize
\texttt{\arxivno{hep-th/0409147}}\\
\texttt{AEI 2004-069}
\end{flushright}
\vspace{2cm}

\begin{center}
{\Large\textbf{\mathversion{bold}Higher-Loop Integrability\\in $\superN=4$ Gauge Theory}\par} \vspace{2cm}

\textsc{Niklas Beisert} \vspace{5mm}

\textit{Max-Planck-Institut f\"ur Gravitationsphysik\\
Albert-Einstein-Institut\\
Am M\"uhlenberg 1, 14476 Potsdam, Germany} \vspace{3mm}

\texttt{nbeisert@aei.mpg.de}\par\vspace*{2cm}\vspace*{\fill}

\textbf{Abstract}\vspace{7mm}

\begin{minipage}{12.7cm}\small
The dilatation operator measures scaling dimensions of
local operator in a conformal field theory.
Algebraic methods of constructing the dilatation operator
in four-dimensional $\superN=4$ gauge theory are reviewed.
These led to the discovery of novel integrable spin chain models
in the planar limit. Making use of Bethe ans\"atze a superficial
discrepancy in the AdS/CFT correspondence was found, we 
discuss this issue and give a possible resolution.
This is the transcript of a talk given at Strings 2004 in Paris.
\end{minipage}\vspace*{\fill}

\end{center}

\newpage

\else

\begin{frontmatter}

\selectlanguage{english}
\title{Higher-Loop Integrability in $\superN=4$ Gauge Theory}

\vspace{-2.6cm}

\selectlanguage{francais}
\title{Here is the French title}

\selectlanguage{english}
\author{Niklas Beisert}
\ead{nbeisert@aei.mpg.de}

\address{Max-Planck-Institut f\"ur Gravitationsphysik,\\
Albert-Einstein-Institut,\\Am M\"uhlenberg 1, 14476 Potsdam, Germany}

\begin{abstract}
The dilatation operator measures scaling dimensions of
local operator in a conformal field theory.
Algebraic methods of constructing the dilatation operator
in four-dimensional $\superN=4$ gauge theory are reviewed.
These led to the discovery of novel integrable spin chain models
in the planar limit. Making use of Bethe ans\"atze a superficial
discrepancy in the AdS/CFT correspondence was found, we 
discuss this issue and give a possible resolution.
This is the transcript of a talk given at Strings 2004 in Paris.

\vskip 0.5\baselineskip

\selectlanguage{francais}
\noindent{\bf R\'esum\'e}
\vskip 0.5\baselineskip
\noindent

\selectlanguage{english}
\end{abstract}

\begin{keyword}
AdS/CFT correspondence \sep integrable spin chains \sep Bethe ansatz\\
\selectlanguage{francais}\noindent{\it Mots-cl\'es: aksdjfh}
\selectlanguage{english}
\PACS
11.15.-q \sep
11.25.Hf \sep
11.25.Tq \sep
02.30.Ik \sep
75.10.Pq 
\end{keyword}

\end{frontmatter}


\selectlanguage{english}

\fi

\section{Introduction}

\paragraph*{Large Spin Limits of AdS/CFT.}

The AdS/CFT correspondence predicts the agreement of spectra of 
energies $E$, in IIB string theory on $AdS_5\times S^5$ and 
scaling dimensions $D$, in $\superN=4$ gauge theory. 
Unfortunately direct tests of this conjecture are prevented by 
the fact that it is a strong/weak duality.
In the last two years, however, there have been two proposals 
how this problem might be circumvented. For these, one focuses on strings 
with a large spin $J$ on $S^5$. In gauge theory these states correspond to 
long local operators.
The first proposal is the celebrated BMN limit by Berenstein, Maldacena 
and Nastase corresponding to strings on plane waves
\cite{Berenstein:2002jq}. The 
second, proposed by Frolov and Tseytlin, is a semiclassical 
limit of string theory \cite{Frolov:2003qc}. 
In these two proposals an effective coupling constant 
\[\lambda'=\frac{\lambda}{J^2}\qquad
\mbox{or}\qquad
\tilde g=\frac{g}{L}\]
emerges. This may be assumed to be 
small, no matter if $\lambda$ itself is small or large. So we 
might expand in the effective coupling in both theories.
On the one hand, in string theory one finds that one can expand 
in $\lambda'$ and $1/J$ which effectively counts sigma-model loops. 
On the other hand, in gauge theory one finds that the $\ell$-loop 
contribution is suppressed by at least $2\ell$ powers of $1/J$. Thus 
we can reorganize the series in powers of $\lambda'$.
So naively the string theory expansion in $\lambda'$ is 
equivalent to the loop expansion in gauge theory and we can go 
ahead and compare.

\paragraph*{Three-Loop Discrepancies.}

Consider a BMN state with two excitations
\[
\Op_n\approx \sum_{p=0}^J \exp \frac{2\pi i np}{J}\Tr \fldZ^p \phi \fldZ^{J-p} \phi
\approx 
\alpha_{+n}^\dagger
\alpha_{-n}^\dagger
\state{0;J},
\qquad
D-J\approx 2\sqrt{1+\frac{\lambda n^2}{J^2}}\,.
\]
In gauge theory we have calculated its dimension up to three loops and 
first order in $1/J$, i.e.~in near BMN limit \cite{Beisert:2003tq}
\[
D-J=2+\frac{\lambda n^2}{J^2}\lrbrk{1-\frac{2}{J}}
-\frac{\lambda^2 n^4}{J^4}\lrbrk{\frac{1}{4}+\frac{0}{J}}
+\frac{\lambda^3 n^6}{J^6}\lrbrk{\frac{1}{8}+\frac{1}{2J}}
+\ldots\,.
\]
In string theory on near plane-waves Callan, Lee, McLoughlin, Schwarz, 
Swanson and Wu have computed the energy to this accuracy and found a 
very similar expression \cite{Callan:2003xr}
\[
E-J=2+\lambda' n^2\lrbrk{1-\frac{2}{J}}
-\lambda'^2 n^4\lrbrk{\frac{1}{4}+\frac{0}{J}}
+\lambda'^3 n^6\lrbrk{\frac{1}{8}+\frac{0}{J}}
+\ldots\,.
\]
However, these two results are not quite identical. The three-loop, $1/J$ 
correction has a different coefficient.
This was only the first sign of a disagreement, it can be observed for 
three excitation BMN operators as well \cite{Callan:2004ev}. 
Even more interestingly, Serban and Staudacher discovered a way to 
compute the three-loop dimensions for states dual to semiclassical 
spinning strings \cite{Serban:2004jf}. 
Again they found a mismatch starting only at three-loops, 
here it is not merely a disagreement of coefficients, but rather 
a disagreement on a functional level. So we see that there is a genuine problem here.

\paragraph*{Overview.}

In the following, I would like to focus on how to obtain these results 
in gauge theory. 
Here, integrability plays a major role, especially at higher-loops. 
I will thus explain this feature and describe how one can make use of it.
Finally, I would like to reconsider the discrepancy and comment 
on the impact on the AdS/CFT correspondence.
Throughout the talk, I will consider scaling dimensions of 
local operators in $U(N)$ $\superN=4$ conformal gauge theory. 
\[
\vev{\Op(x)\,\Op(y)}\sim\frac{1}{|x-y|^{2D}}\,.
\]
Here I will restrict to the planar limit.

\section{Dilatation Operator}

\paragraph*{Single Trace Operators and Spin Chains.}

\begin{figure}
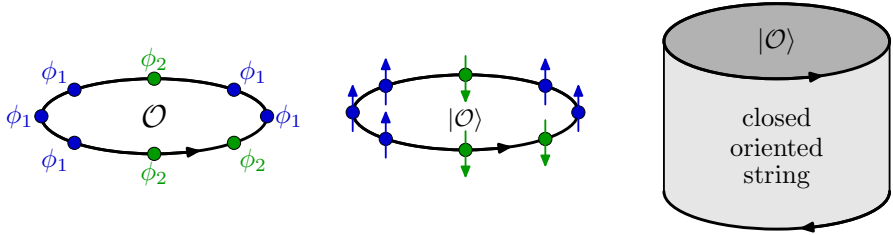
\centering
\parbox{4cm}{\centering\includegraphics[scale=1.0]{SemStrings04.SpinChain.Trace.eps}}
\parbox{4cm}{\centering\includegraphics[scale=1.0]{SemStrings04.SpinChain.Spins.eps}}
\parbox{4cm}{\centering\includegraphics[scale=1.0]{SemStrings04.SpinChain.String.eps}}
\caption{Duality between gauge theory local operators and spin chains 
and similarity to a closed string.}
\label{fig:1}
\end{figure}

Let me briefly review the duality between single-trace operators and 
spin chains without going into details.
Let me first concentrate on two complex scalars $\phi_1$ and $\phi_2$. 
These are also known as $(\fldZ,\phi)$ or $(\fldZ,\fldX)$.
To construct local operators one takes a trace of a product of these 
fields
\[\Op=\Tr \phi_1\phi_1\phi_2\phi_1\phi_1\phi_1\phi_2\phi_2.\]
We now identify $\phi_1$ with spin up and $\phi_2$ with spin down. The state 
can be written as a spin chain state
\[\state{\Op}=\state{
\uparrow\uparrow\downarrow\uparrow\uparrow\uparrow\downarrow\downarrow}.
\]
The number of spin sites or equivalently the number of fields will be 
called the length $L$ of the state.
Note that there is operator mixing, i.e.~one has to deal with linear 
combinations of these pure states
\[\Op=\ast\,\state{\ldots}+\ast\,\state{\ldots}+\ldots\,.\]
Note also that the trace or the spin chain are to be interpreted as a 
closed string in the AdS/CFT correspondence
(see Fig.~\ref{fig:1}).

\paragraph*{Full $\superN=4$ SYM and Subsectors.}

\begin{figure}\centering
\includegraphics[scale=1.0]{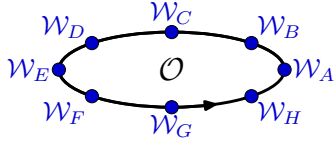}
\caption{Generic Spin Chain.}
\label{fig:2}
\end{figure}

\begin{figure}
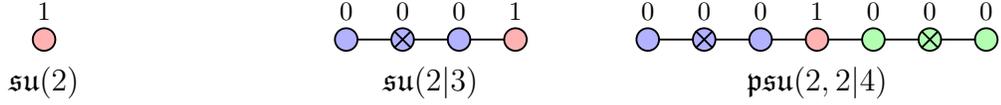
\centering
\parbox{5cm}{\centering\includegraphics[scale=1.0]{SemStrings04.Dynkin.SU2.eps}\\$\alSU(2)$}
\parbox{5cm}{\centering\includegraphics[scale=1.0]{SemStrings04.Dynkin.SU23.eps}\\$\alSU(2|3)$}
\parbox{5cm}{\centering\includegraphics[scale=1.0]{SemStrings04.Dynkin.SU224.eps}\\$\alPSU(2,2|4)$}
\caption{Dynkin diagrams and Dynkin labels for subsectors.}
\label{fig:3}
\end{figure}

This is certainly not the full story because one can put any of the 
fields at the spin sites, not just $\phi_1$ and $\phi_2$
(see Fig.~\ref{fig:2})
\[\Op=\Tr \fld_A\fld_B\fld_C\fld_D\fld_E\fld_F\fld_G\fld_H.
\]%
When we restrict to these two we get the $\alSU(2)$ subsector. It is called 
$\alSU(2)$ because the two fields 
$\fld_A\in\set{\phi_{1},\phi_{2}}$ transform in the fundamental 
representation of $\alSU(2)$.
Another interesting subsector, which will be used later, consists of three 
scalars and two fermions, $\fld_A\in\set{\phi_{1,2,3},\psi_{1,2}}$. 
They transform in the fundamental 
representation of $\alSU(2|3)$, so in fact this subsector is supersymmetric.
For the full $\superN=4$ gauge theory we may in addition use field strengths 
and covariant derivatives,
$\fld_A\in\set{\cder^k\Phi,\cder^k\Psi,\cder^k\fldF}$. 
Note that the derivatives do not constitute 
independent spin sites, they are always associated to the scalars, 
fermions or field strengths.
As we can put arbitrarily many covariant derivatives at each site, we 
have an infinite-dimensional or non-compact representation of the 
symmetry group, which is the $\superN=4$ superconformal group $\alPSU(2,2|4)$. 
This might be somewhat scary at first sight, but it will turn out not to be 
such a big difference (see Fig.~\ref{fig:3}).
There are many more interesting subsectors, a 
classification can be found in \cite{Beisert:2004ry}.

\paragraph*{Dilatation Generator.}

\begin{figure}
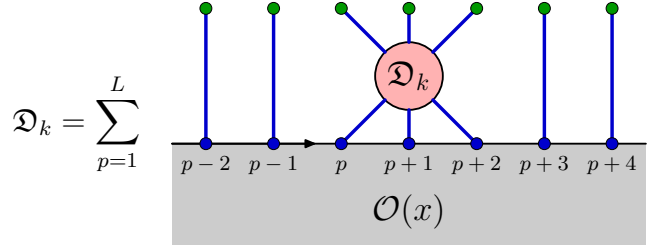
\centering
$\displaystyle
\dil_{k}
=
\sum_{p=1}^L
\ifarxiv 
\parbox{7.0cm}{\centering\includegraphics[scale=0.9]{SemStrings04.Action.eps}}
\else
\parbox{5.5cm}{\centering\includegraphics[scale=0.7]{SemStrings04.Action.eps}}
\fi
$
\caption{Action of the dilatation operator.}
\label{fig:4}
\end{figure}

For the comparison with string theory, we would like to compute scaling 
dimensions. They can be conveniently obtained as the eigenvalues of the 
dilatation operator, we simply have to solve the eigenvalue problem
\[
\dil(g)\Op=D_{\Op}(g)\Op.
\]
The dilatation operator can be computed in perturbation theory
in $g\sim\sqrt{\lambda}$, where we know how to handle gauge theory. 
There is a classical piece $\dil_0$, a one-loop piece $\dil_2$ and higher-loop pieces
$\dil_{3,4,\ldots}$
\[
\dil(g)=\dil_0+g^2\dil_2+g^3\dil_3+g^4\dil_4+\ldots\,.
\]
In the planar limit, the contributions to the dilatation operator act 
locally and homogeneously along the spin chain. They take a few 
adjacent fields and transform them into some other fields
(see Fig.~\ref{fig:4})
\[
\dil_k=\sum_{p=1}^L \dil_{k,p}.
\]

\paragraph*{One-Loop.}

\begin{figure}
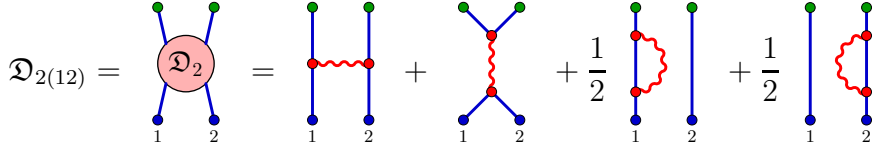
\centering
$\displaystyle\dil_{2(12)}=
\parbox[c]{1.5cm}{\centering\includegraphics[scale=0.75]{SemStrings04.One.Sum.eps}}
=
\parbox[c]{1.5cm}{\centering\includegraphics[scale=0.75]{SemStrings04.One.A.eps}}
+\parbox[c]{1.5cm}{\centering\includegraphics[scale=0.75]{SemStrings04.One.B.eps}}
+\frac{1}{2}\parbox[c]{1.5cm}{\centering\includegraphics[scale=0.75]{SemStrings04.One.C.eps}}
+\frac{1}{2}\parbox[c]{1.5cm}{\centering\includegraphics[scale=0.75]{SemStrings04.One.D.eps}}
$
\caption{One-loop contributions to the dilatation operator.}
\label{fig:5}
\end{figure}

The one-loop correction to the dilatation operator takes two fields 
into two fields. Here, few types of Feynman diagrams 
contribute and the dilatation operator is determined by their 
logarithmic pieces (see Fig.~\ref{fig:5})
First of all for the $\alSU(2)$ subsector we get after integration 
simply ``identity minus permutation'', i.e.~do not modify the two involved 
fields minus interchange the two fields 
\cite{Minahan:2002ve}
\[
\dil_{2(12)}=1-P_{(12)}.
\]
This is exactly isomorphic to 
the so-called Heisenberg XXX$_{1/2}$ spin chain Hamiltonian.
The generalization to the supersymmetric subsector is straightforward. 
Simply replace the permutation by a graded permutation to account for 
the presence of fermions \cite{Beisert:2003ys}
\[
\dil_{2(12)}=1-SP_{(12)}.
\]
For the full $\superN=4$ theory it is a bit more involved. The action of the 
dilatation operator on two fields is given by the harmonic series up to 
their total spin \cite{Beisert:2003jj}
\[
\dil_{2(12)}=2h(J_{(12)}),\qquad
h(s)=\sum_{k=1}^s\frac{1}{k}\,.
\]
The total spin is a superconformal invariant of two 
fields in analogy with the total angular momentum of two spins of the 
rotation group.

\paragraph*{Higher-Loops.}

\begin{figure}\centering
$\dil_{3}=
\parbox[c]{2cm}{\centering\includegraphics[scale=0.7]{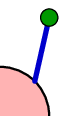}}
+\parbox[c]{2cm}{\centering\includegraphics[scale=0.7]{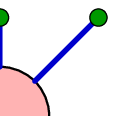}},
$%
\ifarxiv
\par\vspace{0.4cm}
\else
\qquad
\fi%
$
\dil_{4}=
\parbox[c]{2.5cm}{\centering\includegraphics[scale=0.7]{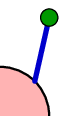}}
+\parbox[c]{2cm}{\centering\includegraphics[scale=0.7]{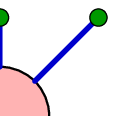}}
+\parbox[c]{2.5cm}{\centering\includegraphics[scale=0.7]{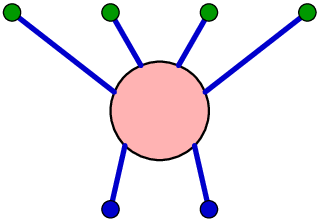}}
+\ldots
$
\caption{Higher-loop contributions to the dilatation generator.}
\label{fig:6}
\end{figure}

\begin{figure}\centering
$
\gen[Q]_{1},\gen[P]_{1},\gen[S]_{1},\gen[K]_{1}=
\parbox[c]{1.5cm}{\centering\includegraphics[scale=0.7]{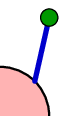}}
+
\parbox[c]{1.5cm}{\centering\includegraphics[scale=0.7]{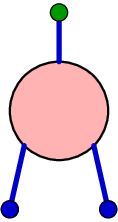}}
$
\caption{Corrections to the (super)momenta and (super)boosts.}
\label{fig:7}
\end{figure}

The one-loop contribution involves four fields, two incoming and two outgoing 
ones. By inspecting Feynman diagrams it is a straightforward to show 
that the order $g^k$ contribution has no more than $k+2$ legs.
So at third order in $g$ we have five legs and so on (see Fig.~\ref{fig:6}).
An exciting feature of higher loops is that now the number of spin 
sites can fluctuate and a novel kind of spin chain emerges
\cite{Beisert:2003ys}.
Note also that at higher loops one has to take into account that the 
order generators of the superconformal algebra are corrected
(see Fig.~\ref{fig:7})
%

\paragraph*{Algebraic Construction.}

In principle we could now go to higher-loops and compute the 
contributions to the dilatation operator. A direct calculation however 
becomes very hard very soon.
The alternative that I propose is to try to reconstruct the dilatation 
operator from some of its known properties.

First of all one needs to consider all possible independent structures, 
which are mainly restricted by the number of fields that contribute. 
For higher-loop contributions we might easily have hundreds of structures.
Then one assumes the most generic form by taking a linear combination 
of the structures with undetermined coefficients.
Finally, one demands the closure of the symmetry algebra, e.g.
\[
\comm{\dil(g)}{\gen[Q](g)}=+\half\, \gen[Q](g),\quad
\comm{\dil(g)}{\gen[S](g)}=-\half\, \gen[S](g),\quad \ldots\,.
\]
This means that the algebra relations must be satisfied exactly 
or, at least, in perturbation theory which is what we are interested in. This 
gives tight constraints and usually only a few coefficients remain 
undetermined.

To fix the remaining ones, we can make use of further constraints.
First of all, we might use BMN scaling behavior, i.e.~that the 
dimension admits an expansion in powers of $\lambda'$.
Then we can make use of known scaling dimensions, for example the one 
of the Konishi operator at one loop.
Finally, we might use the assumption of integrability. I will come to 
this point below.

\paragraph*{Algebraic Construction: Results.}

Using these proposed methods, I have managed to compute the full 
one-loop dilatation operator of $\superN=4$ SYM. This allows to compute any 
one-loop scaling dimension purely algebraically without having to deal 
with integrals or divergencies. What is remarkable about this is that 
the superconformal algebra completely fixes the dilatation operator up 
to one overall constant, which is the coupling constant $g$
\cite{Beisert:2004ry}. 

To proceed to higher-loops it is convenient to restrict to a subsector 
to reduce the complexity of the calculations. For example take the 
supersymmetric $\alSU(2|3)$ subsector. There, I was able to obtain a 
unique result by assuming symmetry and BMN scaling \cite{Beisert:2003ys}.
We can now evaluate the dimension of the Konishi operator up to three 
loops confirming an earlier conjecture in \cite{Beisert:2003tq}
\[
D_{\mathcal{K}}=2+\frac{3\lambda}{4\pi^2}
-\frac{3\lambda^2}{16\pi^4}
+\frac{21\lambda^3}{256\pi^6}+\ldots\,.
\]
The two-loop result was already known and agrees with our 
computation. Our three-loop coefficient has just recently been 
confirmed by a direction computation within QCD which involved of the 
order of 100000 integrals to be computed 
\cite{Moch:2004pa} and also some educated guessing 
of how to extend the QCD result to $\superN=4$ gauge theory
\cite{Kotikov:2004er}.
As an aside, let me note that the dilatation operator exactly agrees 
with the BMN matrix model \cite{Klose:2003qc}. 
In this context, the BMN matrix model is a 
theory quite similar to $\superN=4$ SYM and apparently it might even agree 
with it in this subsector.
Finally, let me also note that this dilatation operator yields the near 
BMN result from the beginning which does not agree with string theory.

When we restrict even further, we find that we can obtain a unique 
result up to at least five loops, if we use the assumption of 
integrability \cite{Beisert:2004hm}.

\section{Integrability}

\paragraph*{One-Loop Integrability.}

Now let us consider integrability. All above computations could in 
principle be generalized to $1/N$ non-planar corrections. For 
integrability, however, we have to consider the strict planar limit.
Integrability now means that next to the dilatation operator there 
exist charges $\charge_{2,3,4}$ of a similar form. These commute with the symmetry 
algebra and among themselves
\[
\comm{\genJ_0}{\charge_r}=\comm{\charge_r}{\charge_s}=0.
\]
The second charge is given precisely by the dilatation operator, the 
third one by a particular combination and so on (see Fig.~\ref{fig:8})
\[
\dil_2=\charge_2,
\qquad
\charge_{3,123}=\sfrac{i}{2}\comm{\dil_{2,12}}{\dil_{2,23}}.
\]%
\begin{figure}\centering
$\displaystyle
\parbox{1.5cm}{\centering\includegraphics[scale=0.7]{SemStrings04.Charge.2.eps}}
=
\parbox{1.5cm}{\centering\includegraphics[scale=0.7]{SemStrings04.Charge.2A.eps}}
\qquad
\qquad
\parbox{1.5cm}{\centering\includegraphics[scale=0.5]{SemStrings04.Charge.3.eps}}
=
\frac{i}{2}\parbox{2cm}{\centering\includegraphics[scale=0.5]{SemStrings04.Charge.3A.eps}}
-\frac{i}{2}\parbox{2cm}{\centering\includegraphics[scale=0.5]{SemStrings04.Charge.3B.eps}}
\qquad
\qquad
\ldots
$
\caption{The charges $\charge_2$ and $\charge_3$.}
\label{fig:8}
\end{figure}%
These charges commute, 
which requires $\dil_2$ to be of a very spacial form, i.e.~integrable.

Minahan and Zarembo found that the planar one-loop dilatation operator 
in the $\alSO(6)$ subsector is indeed integrable \cite{Minahan:2002ve}. 
Let me note that integrability has been found several years ago in 
some subsectors of large $N_c$ QCD 
(see e.g.~\cite{Belitsky:2004cz} for a review).
We have then generalized these results to the full $\superN=4$ theory. 
There we get a super spin chain with $\alSU(2,2|4)$ symmetry \cite{Beisert:2003yb}.

\paragraph*{Test for Integrability.}

\begin{figure}\centering
$
\parbox{5.0cm}{\centering\includegraphics[scale=1.0]{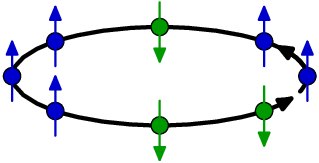}}
\stackrel{\parity}{\longleftrightarrow}
\parbox{5.0cm}{\centering\includegraphics[scale=1.0]{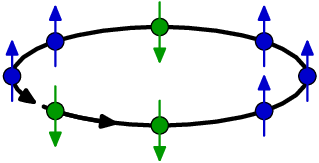}}
$
\caption{Parity operation.}
\label{fig:9}
\end{figure}

To prove integrability is not very easy, but there is a quick and 
reliable test. It involves a parity operator $\parity$ 
which inverts the order of fields within the trace 
or equivalently which flips the spin chain
(see Fig.~\ref{fig:9})
\[
\Tr {\phi_1}{\phi_1}{\phi_2}{\phi_1}{\phi_1}{\phi_1}{\phi_2}{\phi_2}
\stackrel{\parity}{\longleftrightarrow}
\Tr {\phi_2}{\phi_2}{\phi_1}{\phi_1}{\phi_1}{\phi_2}{\phi_1}{\phi_1}.
\]%
It turns out that the even integrable charges have even parity while 
the odd ones have odd parity
\[\parity \charge_r \parity^{-1} = (-1)^r \,\charge_r.\]
The commuting of the charges now leads to a paired spectrum. Paired 
means that for almost every state there is another state with opposite 
parity and exactly degenerate dimension
\[D_+=D_-.\]
Let me note that this is so 
only in the planar limit, $1/N$ corrections destroy these degeneracies, 
so integrability is a genuinely planar effect.
Pairing of states appears to be quite a reliable test for integrability
\cite{Grabowski:1995rb}.

\paragraph*{Higher-Loop Integrability.}

Up to this point we have discussed integrability only at the one-loop level. 
Unfortunately, the structure of the higher-loop corrections seems to 
prevent to employ the usual R-matrix formalism or to prove a 
Yang-Baxter equation.
However, we can rely on our earlier definition of integrability 
involving the higher charges. Higher-loop integrability means that 
there exist interacting charges $\charge_r(g)$, 
i.e.~charges which depend on the 
coupling constant, which commute with the interacting symmetry algebra 
and among themselves \cite{Beisert:2003tq}
\[
\comm{\genJ(g)}{\charge_r(g)}=\comm{\charge_r(g)}{\charge_s(g)}=0.
\]
Again, the dilatation operator is related to the 
second charge via
\[\dil(g)=\dil_0+g^2\charge_2(g).\]
In practice it is easy construct the interacting charges. We can 
however make use of the test from above. It 
requires that the higher-loop corrections to the scaling dimensions do 
not break the degeneracy of pairs
\[
D_+(g)=D_-(g).
\]

Here we cannot consider the $\alSO(6)$ subsector for which 
integrability was originally found because it is not closed at 
higher-loops due to mixing with fermions.
Instead we can consider the $\alSU(2|3)$ sector, there I found that the 
three-loop dilatation operator is integrable by means of these pairs
\cite{Beisert:2003ys}. 
This is very interesting, because here the length of the spin chain is 
allowed to fluctuate. This might seem to violate the prerequisites for 
integrability, but apparently it works out. 
We can also make good use of the integrability. For instance we found 
that the BMN limit together integrability determines (at least) the 
five-loop contribution to the dilatation operator uniquely
\cite{Beisert:2004hm}. What is 
remarkable is that we find the quantitative BMN square-root formula
\[
D-J=\sum_k\sqrt{1+\lambda'n_k^2}
\]
just by assuming the qualitative BMN limit.

\paragraph*{Long-Range Bethe Ansatz.}

Integrability offers a very powerful tool to compute planar scaling 
dimensions, the so-called algebraic Bethe ansatz. 
Serban and Staudacher have found out that the $\alSU(2)$ subsector is 
isomorphic to the Inozemtsev spin chain up to three loops \cite{Serban:2004jf}. 
Fortunately, 
the Bethe equations are known for this model and allow to compute 
three-loop scaling dimensions. We have then figured out how to modify 
the equations to account for presumably all-loop effects in an 
asympotical sense \cite{Beisert:2004hm}.

Our proposal is that this set of algebraic equations for the Bethe 
roots $u_k$ describes planar anomalous dimensions of states in the 
$\alSU(2)$ up to very high loop orders. 
\[
\frac{x(u_k-\sfrac{i}{2})^L}{x(u_k+\sfrac{i}{2})^L}=
\prod_{j=1}^K \frac{u_k-u_j-i}{u_k-u_j+i}\,,
\qquad
x(u)=\frac{u}{2}+\frac{u}{2}\sqrt{1-\frac{2g^2}{u^2}}\,.
\]
with 
\[
Q_r=\sum_{k=1}^K \frac{i}{r-1}\lrbrk{\frac{1}{x(u_k-\sfrac{i}{2})^{r-1}}-\frac{1}{x(u_k+\sfrac{i}{2})^{r-1}}}
,\qquad
D=L+g^2Q_2.
\]
Asymptotically means that the loop order $\ell$ is only limited by the 
length $L$ of the spin chain. This appears to be no restriction because 
we shall consider very long spin chains to make contact with strings.
When we set the coupling constant to zero, the function $x$ becomes the 
identical function. We then recover the usual one-loop Bethe equations 
for the Heisenberg XXX$_{1/2}$ model.
So far the equations are merely a conjecture, but we have shown 
agreement with the five-loop model for all states of length up to 8. 
It is therefore pretty sure that the Bethe ansatz coincides with our 
spin-chain model.

\section{Discrepancies}

\paragraph*{Spinning Strings at Higher-Loops.}

To make contact with spinning strings one considers the thermodynamic 
limit of long spin chains with a large number of excitations. Here one 
gets an effective coupling $\tilde g=g/L$ in analogy to the BMN effective 
coupling $\lambda'$. 
The Bethe roots now condense on cuts $\contour$ in the complex plane 
with a density function $\rho$. 
The Bethe equation becomes an integral equation 
\[
\frac{1}{\tilde x}
+2\pi n_{\tilde x}\lrbrk{1-\frac{\tilde g^2}{2\tilde x^2}}
=
\pint_{\contour}d\tilde x'\rho(\tilde x')
\lrbrk{
\frac{2}{\tilde x-\tilde x'}
+\frac{\tilde g^2}{\tilde x^2\tilde x'}\,
 \frac{1}{1-\frac{\tilde g^2}{2\tilde x'\tilde x}}
}
\]
and the charges, in particular the dimension, are given by 
\[
Q_r=\frac{1}{L^{r-1}}
\int_{\contour}\frac{d\tilde x\,\rho(\tilde x)}{1-\frac{\tilde g^2}{2\tilde x^2}}
\,\,
\frac{1}{\tilde x^r},
\qquad
D=L+g^2Q_2.
\]
Kazakov, Marshakov, Minahan and 
Zarembo have derived Bethe equations for the classical string theory 
\cite{Kazakov:2004qf}. 
One can write them as
\[
\frac{1}{\tilde x}
+2\pi n_{\tilde x}
\lrbrk{1-\frac{\tilde g^2}{2\tilde x^2}}
=
\pint_{\mathcal{C}}d\tilde x'\rho(\tilde x')
\lrbrk{
\frac{2}{\tilde x-\tilde x'}
+\frac{\tilde g^2}{\tilde x^2\tilde x'}\,
\frac{1}{1-\frac{\tilde g^2}{2\tilde x'\tilde x'}}
},
\]
where remarkably the expressions for the charges 
are identical to the expressions in gauge theory \cite{Beisert:2004hm}. 
The equations are in 
fact nearly the same, they only differ by the little prime at the last $\tilde x$.
This makes the agreement at two-loops and the disagreement at 
three-loops manifest. The difference between the two equations is 
protected by three powers of $g^2$ (two powers are in the equations, 
one is in the definition of $D$ in terms of $Q_2$).

The string Bethe equations are valid for the classical theory only. It 
would be great if they could be generalized to the quantum model. It 
appears that a first step in this direction was made in 
\cite{Arutyunov:2004vx}.
Their equations not only reproduce the $1/J$ near BMN limit, 
but also a generic $\sqrt[4]{\lambda}$ behavior for large $\lambda$.

\paragraph*{Order of Limits.}

\begin{figure}
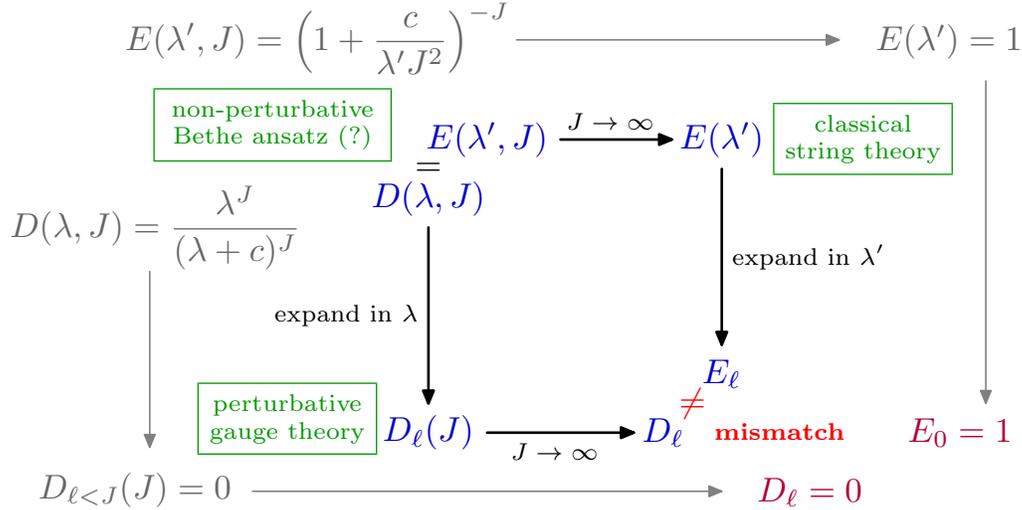
\centering
\ifarxiv
\includegraphics[scale=1.3]{SemStrings04.NonCommute.eps}%
\else
\includegraphics[scale=1.0]{SemStrings04.NonCommute.eps}%
\fi
\caption{Order of limits problem.}
\label{fig:10}
\end{figure}
Now let me try to explain the apparent disagreement at three loops
(see Fig.~\ref{fig:10}).
Assume we have an exact, non-perturbative scaling dimension. We could 
write it as a function $D$ of lambda and $J$ or as a function $E$ of 
$\lambda'$ and $J$, both forms are equivalent. As an example let us take 
\[
D(\lambda,J)=\frac{\lambda^J}{(\lambda+c)^J}\,.
\]
This function is chosen on purpose, it is proportional 
to $\lambda^J$, which will become important below. The function is 
equivalent to 
\[
E(\lambda',J)=\lrbrk{1+\frac{c}{\lambda'J^2}}^{-J}
\]
upon identification of $\lambda'$ with $\lambda/J^2$.
In practice we would not be able to compute this function, but we can 
hope that one day this might be possible, perhaps with a non-perturbative Bethe 
ansatz.

In gauge theory we can do perturbation theory for small $\lambda$. 
In our example we would find that the first few loop contributions vanish.
In string theory we can only access the classical regime. In order for 
this to make sense, $\lambda$ must be large or equivalently the spin $J$
must be large. In our example we find a constant limiting function 
which is simply $1$.

We certainly cannot compare these two results, but BMN and FT proposed 
to consider both limits at the same time and then compare. In gauge 
theory consider large spin $J$ and in string theory consider small 
$\lambda'$.
When we do this in our example we find that obviously all $D_\ell$ are 
zero while $E_0$ equals $1$, although we started off with the same function. 
How can this be? Well, this is a classical order of limits problem, 
there is actually no reason why the two should agree! 
We see that we cannot in fact compare in perturbation theory. Therefore 
the spinning strings and near BMN proposals do not have to work and 
indeed we find a disagreement, but only at higher-loop orders
(the fact that the strict BMN proposal works is 
related to the matching of free magnon energies 
\cite{Beisert:2004hm}).
This is unfortunate. Nevertheless we see that integrability has helped 
in obtaining very precise results in both theories. We may hope that if 
we make full use of it, we could finally solve either theory and thus 
see whether scaling dimensions truly agree with string energies. After 
all, the AdS/CFT correspondence might be valid only approximately and 
truly break down at three loops. Certainly, we currently cannot say at 
the moment.

\paragraph*{Wrapping Interactions.}

\begin{figure}
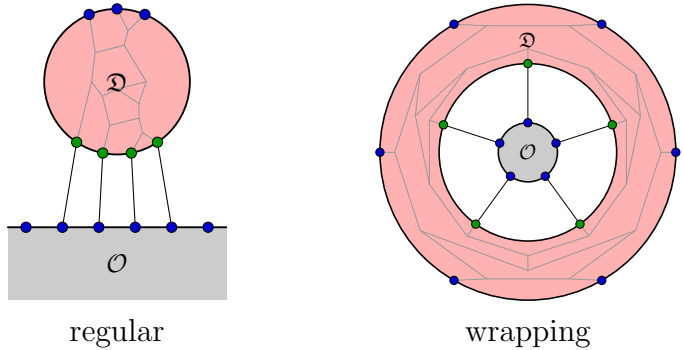
\centering
\parbox[c]{4.5cm}{\centering\includegraphics[height=4cm]{SemStrings04.Type.Regular.eps}\\
regular}
\qquad\parbox[c]{4.5cm}{\centering\includegraphics[height=4cm]{SemStrings04.Type.Wrapping.eps}\\
wrapping}
\caption{Regular and wrapping interactions.}
\label{fig:11}
\end{figure}

Let me just propose a possible and more explicit explanation for the 
discrepancy.
So far we have focussed on planar interactions of regular type which 
merely attach to the state (see Fig.~\ref{fig:11}).
But at very high loop orders there is 
another type of planar diagram which completely wraps the state. 
They start to contribute at $L$ loops where $L$ is the length of the state. 
Naively they would not contribute for very long states, but as we have 
seen in the example, this is not necessarily true. These contributions 
may account for the discrepancy.
Note that the asymptotic Bethe ansatz does not incorporate wrappings, 
it is valid only to the order where they start to contribute. We hope 
to obtain equations to take care of the wrappings some day.
Unfortunately, it seems that the algebraic methods explained at the 
beginning of the talk are not effective for wrappings, so we cannot say 
much at the moment.

\section{Conclusions}

In the above, I have presented methods to conveniently obtain higher-loop 
scaling dimensions of local operators in gauge theory. I have explained 
higher-loop integrability and tried to convince the reader 
that $\superN=4$ SYM has this feature. 
I have then proposed a set of Bethe equations that allow 
to compute planar anomalous dimensions to all-loops in an asymptotic 
sense. The most important open problem here is to find a truly 
non-perturbative extension of these equations, if this should be
possible at all.
Using these methods we were able to detect discrepancies between string 
theory and gauge theory starting at third order in $\lambda$. I have 
argued that these might be due to an order of limits issue and we 
cannot in fact avoid the strong/weak duality by the near BMN and 
Frolov-Tseytlin proposals. An obvious question is now if the non-planar 
extension to the BMN proposal suffers from the same problems. So far it 
has only been confirmed at one-loop and even there there are many 
questions left unanswered, see e.g.~\cite{Gutjahr:2004dv}.

There are many other important things to be done, let me name a few.
First of all one could try to generalize the higher-loop results to 
larger subsectors or even the full theory.
So far we have merely observed integrability by constructing explicitly 
the dilatation generator. Can we somehow prove it from field theory 
arguments? If so, we could maybe use it to go to even higher loops.
Along the same lines it would be important to show integrability for 
the quantum string sigma model. For the classical one the 
existence of a family of flat connections was shown in \cite{Bena:2003wd}.
Finally there is an issue that intrigues me. On the one hand, I have used 
conformal symmetry to construct the dilatation operator at one-loop and 
found it to be integrable. On the other hand, one could demand 
integrability and find exactly the same result which is then 
conformally symmetric. So in some sense, conformal symmetry and 
integrability go hand in hand here, 
even through this not a two-dimensional theory, but
a four-dimensional one. 
I would really like to understand this point better.

\section*{Acknowledgements}

I thank Virginia Dippel, Charlotte Kristjansen and Matthias Staudacher
for their collaboration on the projects on which this text is based.

\ifarxiv\newpage\fi

\bibliographystyle{nbshort}
\bibliography{Strings04Proceed}

\end{document}